\documentstyle[12pt]{article}

\catcode`\@=11
\long\def\@makefntext#1{
\protect\noindent \hbox to 3.2pt {\hskip-.9pt
$^{{\ninerm\@thefnmark}}$\hfil}#1\hfill}		%CAN BE USED

 \def\@makefnmark{\hbox to 0pt{$^{\@thefnmark}$\hss}}  %ORIGINAL

\def\ps@myheadings{\let\@mkboth\@gobbletwo
\def\@oddhead{\hbox{}
\rightmark\hfil\ninerm\thepage}
\def\@oddfoot{}\def\@evenhead{\ninerm\thepage\hfil
\leftmark\hbox{}}\def\@evenfoot{}
\def\sectionmark##1{}\def\subsectionmark##1{}}

\newcounter{sectionc}\newcounter{subsectionc}\newcounter{subsubsectionc}
\renewcommand{\section}[1] {\vspace{0.6cm}\addtocounter{sectionc}{1}
\setcounter{subsectionc}{0}\setcounter{subsubsectionc}{0}\noindent
	{\bf\thesectionc. #1}\par\vspace{0.4cm}}
\renewcommand{\subsection}[1] {\vspace{0.6cm}\addtocounter{subsectionc}{1}
	\setcounter{subsubsectionc}{0}\noindent
	{\it\thesectionc.\thesubsectionc. #1}\par\vspace{0.4cm}}
\renewcommand{\subsubsection}[1]
{\vspace{0.6cm}\addtocounter{subsubsectionc}{1}
	\noindent {\rm\thesectionc.\thesubsectionc.\thesubsubsectionc.
	#1}\par\vspace{0.4cm}}

\newcounter{appendixc}
\newcounter{subappendixc}[appendixc]
\newcounter{subsubappendixc}[subappendixc]

\renewcommand{\appendix}[1] {\vspace{0.6cm}
        \refstepcounter{appendixc}
        \setcounter{figure}{0}
        \setcounter{table}{0}
        \setcounter{equation}{0}
        \renewcommand{\thefigure}{\Alph{appendixc}.\arabic{figure}}
        \renewcommand{\thetable}{\Alph{appendixc}.\arabic{table}}
        \renewcommand{\theappendixc}{\Alph{appendixc}}
        \renewcommand{\theequation}{\Alph{appendixc}.\arabic{equation}}
        \noindent{\bf Appendix \theappendixc #1}\par\vspace{0.4cm}}

\def\abstracts#1{{
	\centering{\begin{minipage}{30pc}\tenrm\baselineskip=12pt\noindent
	\centerline{\tenrm ABSTRACT}\vspace{0.3cm}
	\parindent=0pt #1
	\end{minipage}}\par}}

\renewenvironment{thebibliography}[1]
	{\begin{list}{\arabic{enumi}.}
	{\usecounter{enumi}\setlength{\parsep}{0pt}
\setlength{\leftmargin 1.25cm}{\rightmargin 0pt}
	 \setlength{\itemsep}{0pt} \settowidth
	{\labelwidth}{#1.}\sloppy}}{\end{list}}

\newcounter{itemlistc}
\newcounter{romanlistc}
\newcounter{alphlistc}
\newcounter{arabiclistc}

\newcommand{\fcaption}[1]{
        \refstepcounter{figure}
        \setbox\@tempboxa = \hbox{\tenrm Fig.~\thefigure. #1}
        \ifdim \wd\@tempboxa > 6in
           {\begin{center}
        \parbox{6in}{\tenrm\baselineskip=12pt Fig.~\thefigure. #1}
            \end{center}}
        \else
             {\begin{center}
             {\tenrm Fig.~\thefigure. #1}
              \end{center}}
        \fi}

\newcommand{\tcaption}[1]{
        \refstepcounter{table}
        \setbox\@tempboxa = \hbox{\tenrm Table~\thetable. #1}
        \ifdim \wd\@tempboxa > 6in
           {\begin{center}
        \parbox{6in}{\tenrm\baselineskip=12pt Table~\thetable. #1}
            \end{center}}
        \else
             {\begin{center}
             {\tenrm Table~\thetable. #1}
              \end{center}}
        \fi}

\def\@citex[#1]#2{\if@filesw\immediate\write\@auxout
	{\string\citation{#2}}\fi
\def\@citea{}\@cite{\@for\@citeb:=#2\do
	{\@citea\def\@citea{,}\@ifundefined
	{b@\@citeb}{{\bf ?}\@warning
	{Citation `\@citeb' on page \thepage \space undefined}}
	{\csname b@\@citeb\endcsname}}}{#1}}

\newif\if@cghi
\def\cite{\@cghitrue\@ifnextchar [{\@tempswatrue
	\@citex}{\@tempswafalse\@citex[]}}
\def\citelow{\@cghifalse\@ifnextchar [{\@tempswatrue
	\@citex}{\@tempswafalse\@citex[]}}
\def\@cite#1#2{{$\null^{#1}$\if@tempswa\typeout
	{IJCGA warning: optional citation argument
	ignored: `#2'} \fi}}

\def\fnt#1#2{\footnotetext{\kern-.3em
	{$^{\mbox{\sevenrm #1}}$}{#2}}}
 1
 1
 1

\font\tenbf=cmbx10
\font\tenrm=cmr10
\font\tenit=cmti10

\font\ninerm=cmr9

\begin{document}

\centerline{\tenbf WILSON LOOPS, $q \bar{q}$ and $3 q$ POTENTIALS,}
\baselineskip=22pt
\centerline{\tenbf BETHE--SALPETER EQUATION}
\vspace{0.8cm}
\centerline{\tenrm N. BRAMBILLA and G.M. PROSPERI
\footnote{Presenting author}}
\baselineskip=13pt
\centerline{\tenit Dipartimento di Fisica dell'Universit\`a di Milano
 and I.N.F.N.}
\baselineskip=12pt
\centerline{\tenit Via Celoria 16, 20133 Milano}
\vspace{0.9cm}
\abstracts{The derivation of the $ q \bar q $ and the $ 3q $ potential for two
dynamical quarks in a Wilson--loop context is reviewed. Some improvements are
introduced. Only the usual assumptions in the evaluation of the Wilson loop
integrals and expansions in the quark velocities are required for the result.
It is shown that under the same assumptions
 it is possible to obtain  the relativistic flux--tube lagrangian and
 a $ q \bar{q}$ Bethe--Salpeter equation with a confining kernel for
 spinless quarks.}

\vfil
%\vspace{0.8cm}
\rm\baselineskip=14pt
\section{ Introduction}
%\vspace{-0.7cm}
%\subsection{Typeset Scripts}
%\vspace{-0.35cm}
In this paper first we  review the derivation
of the $ q \bar q $ and
the $ 3q $ semirelativistic potentials for dynamical quarks as has been
given in preceding papers$^{1}$ (for a general review
 on the subject see$^2$) and introduce
 some significant improvements.  Then we show that, under the same
 assumptions and in the case of spinless quarks, a Bethe--Salpeter
 equation with a confining kernel can be obtained.\par
The basic objects considered in the derivation are the appropriate Wilson loop
integrals $ W_{q \bar q} $ and $ W_{3q} $ and the basic assumptions are:
\par
i)  the quantities
 $ i \ln W $ can be expressed as the sum of a short range
contribution $ i \ln W^{SR} $ and a long range one $ i \ln W^{LR} $ ;
\par
ii) the SR--term can be obtained simply from a perturbative expansion
and the LR--term from a strong coupling expansion
(in practice by the area law).
\par
The improvement consists in the fact that
an  ad--hoc explicit instantaneous approximation is no longer required
and only expansions in the quark velocities are used.
Furthermore, the $O (\alpha^{2}_{\rm s}) $
 contribution is explicitly taken into
account in the static part of the potential and it is shown that
a covariant Lorentz gauge as well as the
Coulomb gauge can be used.

As it is  well known,
the arguments in favour of the two assumptions are
 asymptotic freedom and
the observation that the SR--part of the potential vanishes for $ r \rightarrow
\infty $, while the LR--part vanishes for $ r \rightarrow 0 $. Obviously, with
the simple additivity assumption i),
the resulting potential or kernel
 is expected to be inaccurate at intermediate distances;
interferences of the two mechanisms should be important there.
However, no attempt is made in this paper to use a more
 sophisticated approximation
scheme of the type proposed e.g. in Refs.$^{3}$
 ( see also$^4$).

In Sec. 2 we discuss the evaluation of the Wilson loop integrals, in Secs. 3
and  4 we derive the $  q \bar q $  and the $ 3q $
potentials respectively, in Secs. 5 and 6 we sketch the derivation
 of the flux--tube lagrangian and of the Bethe--Salpeter equation.

\section{ Wilson loop integrals}
%\vspace{-0.7cm}
For the $q\overline{q}$ case the basic object is
 \begin{equation}
W_{q\overline{q}} = \frac{1}{3} \left\langle {\rm Tr \   P}
 \exp \left( ig \oint_{\Gamma} dx^{\mu} \
 A_{\mu}(x) \right) \right\rangle \> .
\end{equation}
Here the integration loop $\Gamma$ is assumed to be made by an
arbitrary  world line
$\Gamma_1$ between an
initial position ${\bf y}_1$ at the time $ t_{\rm i} $ and a final one ${\bf
x}_1$ at the time
  $ t_{\rm f} $  for the quark ($t_{\rm i} < t_{\rm f}$), a similar world
line $\Gamma_2$ described
 in the reverse direction from ${\bf x}_2$ at the time $t_{\rm f}$
to ${\bf y}_2$ at the time $t_{\rm i}$ for the antiquark and two straight lines
at fixed times which  connect ${\bf x}_1$ to ${\bf x}_2$, ${\bf y}_2$
to ${\bf y}_1$ and close
the contour. As usual $ A_{\mu} (x) =
\frac{1}{2} {\lambda}_ {a} A_{\mu} ^{a} (x) $,
  P prescribes the ordering
of the color matrices (from right to left) according to the direction fixed
on the loop and the angular brackets denote
 the functional integration.\par
Integrating explicitly the fermion fields, for any functional
 of the gauge field
alone one obtains
\begin{equation}
\langle f[A] \rangle = \frac{\int {\cal D}[A] M_f(A)
f[A] e^{iS[A]}}
{\int {\cal D}[A] M_f(A) e^{iS[A]}} \> ,
\label{eq:due}
\end{equation}
where $ S[A] $ denotes the pure gauge action
plus the gauge--fixing terms and $ M_{\rm f} [A] $ is the fermionic
determinant
\begin{eqnarray}
  M_{\rm f} [A]  & =&   {\rm Det} \prod_{j} [ 1 +
                             i g A (i \partial - m_j )^{-1} ] =
  \sum_{j} [g \int {\rm d}^4 x
 {\rm Tr} (i A(x) S_{\rm F}^{(m_j)}(0) )\nonumber \\
& - &{1 \over 2} g^2
 \int {\rm d}^4 x \int {\rm d}^4 y {\rm Tr}
 ( A(x) i S_{\rm F}^{(m_j)}(x-y) A(y) i S_{\rm F}^{(m_j)}(y-x) )
+ ... ].
\label{eq:tre}
\end{eqnarray}
\indent
Using the above equations and writing the gauge field lagrangian as
the sum of the free and the interaction parts, $ {\cal L}(A)
={\cal L}_0 + {\cal L}_{\rm int} $,
we have the perturbative expansion
\begin{eqnarray}
& & W_{q \bar{q}}^{\rm pert}  = {1 \over 3} \sum_{n=0}^{\infty}
  \sum_{p=0}^{\infty} {1 \over n!p!} \langle  {\rm  Tr} {\rm P}
 (ig \oint
 {\rm d}z^{\mu}
             A_{\mu} )^n (i \int {\rm d}^4 x {\cal L}_{\rm int}(x))^p
 \rangle_0 =
\nonumber \\
& &  ={1 \over 3} \sum_{n=0}^{\infty} \sum_{p=0}^{\infty} {(ig)^n i^p \over p!}
\int {\rm d}^4 x_1 ... \int {\rm d}^4
x_p
 \oint ... \oint_{z_{1} > z_{2} > \dots > z_n} {\rm d}z_{1}^{\mu
_1} ... {\rm d}z_{n}^{\mu_n}\nonumber \\
& & \langle {\rm Tr} [A_{\mu_1}(z_1) \dots
 A_{\mu_n}(z_n)]
 {\cal  L}_{\rm int}(x_1)\dots {\cal L}_{\rm int}(x_p) \rangle_0,
\label{eq:quattro}
\end{eqnarray}
where, due to (\ref{eq:tre}), the single terms must be understood
  as expansions in $g$ in turn.
Then, identifying $ W^{\rm SR} $ with $ W^{\rm pert} $
according to  assumption ii),  we obtain in graphical
 terms
(we omit graphs that are obtained by permutation of  other ones or
 completely cancelled by renormalization)
\vfill\eject
\begin{figure}
\vspace{4 truecm}
\end{figure}
\begin{equation}
                  i \ln W_{q \bar q}^{\rm SR} =
\label{eq:cinque}
\end{equation}
\noindent
 where
  the external circuit stands for the Wilson loop $\Gamma$,
 and the inserted lines for
ordinary free propagators. Notice the term which includes
 a quark-antiquark loop,
which obviously  comes from (\ref{eq:tre}).

The various quantities occurring in (\ref{eq:cinque})
have been extensively studied from the point of view of
re\-nor\-ma\-li\-za\-tion$^{5}$.
 To our knowledge however
no explicit evaluation in closed
form has been given other than in very special cases$^6$.
For the purpose of the derivation of a semirelativistic
potential, an evaluation in terms of
  an expansion in the quark velocities
 shall be sufficient.

Let  $ (z_j^0 = t_j, \   {\bf z}_j = {\bf z}_j(t)) $ be the equation
 for the world
lines  of the quark and the antiquark
 and set $ \dot{z}^{\mu}_j = dz^{\mu}_j /dt = (1,\   \dot{\bf z}_j)
$. The first--order term in
  $\alpha_{\rm s} = g^2/4 \pi $  can be written explicitly as
\begin{equation}
(i \ln W^{\rm SR})^{(1)}_{q \bar{q}}
 ={4 \over 3} g^2 \int_{t_{\rm i}}^{t_{\rm f}}
{\rm d}t_1 \int_{t_{\rm i}}^{t_{\rm f}} {\rm d}t_2 \dot{z}_1^{\mu}(t_1)
 \dot{z}_2^{\nu}(t_2)
D_{\mu \nu}(z_1(t_1) - z_2(t_2))\> ,
\label{eq:sex}
\end{equation}
where the limit for large $ t_{\rm f} - t_{\rm i} $ has been understood and
the contribution from the equal--time lines are
neglected. Performing the change of
variables $ t = {t_1 + t_2 \over 2}$, $ \tau = t_1 - t_2 $,
expanding $ {\bf z}_1 $ and $ {\bf z}_2 $ around $ t $,
\begin{equation}
  {\bf z}_{1}(t_{1}) ={\bf z}_{1}(t) + {1 \over 2} \tau \dot{\bf z}_{1}(t)
    +{1 \over 8} \tau^2 \ddot{\bf z}_{1}(t)+\dots,\quad\qquad
{\bf z}_{2}(t_2) =  {\bf z}_{2}(t) - {1 \over 2} \dot {\bf z}_{2}(t)+
     {1 \over 8} \tau^2 \ddot{\bf z}_{2}(t)- \dots
\end{equation}
 and integrating over $\tau$ (between $- \infty $ and $ + \infty $),
in the  Coulomb gauge we obtain immediately
\begin{equation}
(i \ln W^{\rm SR})^{(1)} = \int_{t_{\rm i}}^{t_{\rm f}}
{\rm d}t \{ -{4 \over 3} {\alpha_{\rm s} \over r}
[1- {1\over 2}( \delta^{hk} + \hat{r}^h
\hat{r}^k )) \dot{z}_{1}^h \dot{z}_{2}^k + \dots] \}\> ,
\label{eq:otto}
\end{equation}
with ${\bf r}={\bf z}_1 -{\bf z}_2$ and $\hat{\bf r}={{\bf r}/r}$.
If we had  worked, e.g., in the Feynman gauge, we would have obtained
\begin{equation}
(i \ln W^{\rm SR})^{(1)}_{q \bar{q}} =
 \int_{t_{\rm i}}^{t_{\rm f}} {\rm d}t
 \{ -{4 \over 3}
 { \alpha_{\rm s} \over r } [1 - \dot{\bf z}_1 \cdot \dot{\bf z}_2 +
   {1 \over 8} ((\dot{\bf z}_1 + \dot{\bf z}_{2})^2 + {\bf r} \cdot
\ddot{\bf r}) -
{1 \over 8 r^2} (({\bf r}\cdot (\dot{\bf z}_1+\dot{\bf z}_2))^2+\dots
] \}
\label{eq:nove}
\end{equation}
from which  (\ref{eq:otto})
 can be recovered by eliminating the acceleration term by
 partial integration. This is a consequence of
  the gauge invariance of the Wilson
integral.\par
   In a similar way, after renormalization, we can obtain for the
$ \alpha_{\rm s}^2 $ term in the static limit
\begin{equation}
(i \ln W^{\rm {SR}})^{(2)}_{ q \bar{q}}
  = \int_{t_{\rm i}}^{t_{\rm f}} {\rm d}t \, \{-
{4\over 3}{ \alpha_{\rm s}^2 \over
   4 \pi} {1 \over r}[({66-4N_{\rm f} \over 3})(\ln \mu r + \gamma)+A]
+\dots\}.
\label{eq:undici}
\end{equation}
In Eq.(\ref{eq:undici}) $\mu$ is the renormalization scale
 and $A$
 is a constant that depends on the renormalization
 convention. In the $\overline{\rm MS}$ scheme  $A
= {5\over 6}({66-4 N_f\over 3})  -8  $.\par
Let us come to the LR part of the Wilson integral.
 We shall make the assumption
\begin{equation}
i \ln W_{q \bar{q}}^{\rm LR}= \sigma S_{\rm min} +{1\over 2} C P \> ,
\label{eq:dodici}
\end{equation}
where $ S_{\rm min} $ denotes the minimal surface enclosed by the loop $ \Gamma
$ and $ P $ its lenght. Eq. (\ref{eq:dodici})
 is suggested by  the pure lattice gauge
 theory
and it  is believed to be true in the so--called quenched approximation,
i.e. when we replace $ M_{\rm f}(A) $ by 1 in (\ref{eq:due}).
 Corrections to  the pure
potential theory (pair creation effects) should be introduced for this fact but
they shall not be considered here.

In more explicit terms (\ref{eq:dodici}) can be written as
\begin{equation}
  i \ln W^{\rm LR}_{q \bar{q}} = \sigma \, {\rm min}
 \int_{t_{\rm i}}^{t_{\rm f}} {\rm d}t
       \int_0^1 {\rm d}s [-({\partial x \over \partial t})^2 ({\partial x \over
       \partial s})^2 + ({\partial x^{\mu} \over \partial t}{\partial x_{\mu}
       \over \partial s})^2]^{1\over 2}
  + {1 \over 2} C \sum_{j=1,2} \int_{t_{\rm i}}^{t_{\rm f}} {\rm d}t
[{\dot{z}_j^{\mu} \dot{z}_{j \mu}}]^{1\over 2},
\label{eq:tredici}
\end{equation}
where the minimum is taken over all surfaces  of
equation $ x^{\rho}=x^{\rho}(t,\; s) $ having $\Gamma$ as contour.
Obviously $ x^0 = t$,
$ {\bf x}(t,\; 1) = {\bf z}_1 (t)$ and  ${\bf x}(t, \; 0)= {\bf z}_2 (t)$.\par
By solving the appropriate Euler equations and expanding
  in the
velocities, we obtain
\begin{equation}
{\bf x}_{\rm min}(t,\; s) = s \,
 {\bf z}_1 (t) + (1-s)\, {\bf z}_2 (t)
    - {1 \over 2} s (1-s) [{\bf \eta} + {1 \over 3} (1+s) {\bf \zeta}] +
\dots
\label{eq:quindici}
\end{equation}

\noindent with
$
  {\bf \eta}  = (\dot{\bf r}\cdot \dot{\bf z}_2 - {\bf r} \cdot
    \ddot{\bf z}_2) {\bf r} + ({\bf r} \cdot \dot{\bf r}) \dot{\bf z}_2
     - 2 ({\bf r} \cdot \dot{\bf z}_2) \dot {\bf r} + r^2 \ddot{\bf z}_2 $
and
${\bf \zeta} =  - ({\bf r} \cdot \dot{\bf r}) \dot{\bf r} + r^2 \ddot{\bf r}
     + (\dot{r}^2 - {\bf r} \cdot \ddot{\bf r}){\bf r}$.\par
Actually it can be checked that the $ {\rm O} (v^2) $ term
 in (\ref{eq:quindici})
 does not contribute
to $ S_{\rm min} $ at order $v^2$ (such a term is however important in
principle for the evaluation of the functional derivatives).
Replacing (\ref{eq:quindici}) in (\ref{eq:tredici}), finally we have
\begin{eqnarray}
& & \quad  i\ln W^{{\rm LR}}_{q\overline{q}} =
\int_{t_{\rm i}}^{t_{\rm f}} dt \   \sigma r \int_0^1 ds \   [1-(s
\dot{{\bf z}}_{1 \rm T} + (1-s)
 \dot{{\bf z}}_{2 \rm T} )^2]^{\frac{1}{2}} +{1\over 2} C
 \sum_{j=1}^2 \int_{t_i}^{t_f}
 ( 1 - \dot{z}_j^h \dot{z}_j^h)^{1\over 2} =
\nonumber\\
& & =  \int_{t_{\rm i}}^{t_{\rm f}} dt \   \sigma r \
 [ 1-\frac{1}{6}
(\dot{{\bf z}}_{1 \rm T}^2
+ \dot{{\bf z}}_{2 \rm T}^2
+ \dot{{\bf z}}_{1 \rm T} \cdot \dot{{\bf z}}_{2 \rm T} )]
+ \dots
+{1\over 2} C
 \sum_{j=1}^2 \int_{t_i}^{t_f}
 ( 1 -{1\over 2} \dot{z}_j^h \dot{z}_j^h+\dots),
\label{eq:dic}
\end{eqnarray}
where $ \dot{\bf z}_{j {\rm T}}$ denotes the {\it transversal part}
 of $\dot{\bf z}_j $,
$ \dot{z}_{j{\rm T}}^h=  (\delta^{hk} -\hat{r}^h \hat{r}^k)
 \dot{z}_j^h$. \par
 In conclusion, we can write
\begin{equation}
i \ln W_{ q \bar{q}} = ( i \ln W_{ q \bar{q}}^{\rm SR})^{(1)}
+ ( i \ln W_{q \bar{q}}^{\rm SR})^{(2)} + \dots
  + i \ln W_{q \bar{q}}^{\rm LR},
\label{eq:dicbis}
\end{equation}
with the various terms as given by (\ref{eq:otto}), (\ref{eq:undici}),
 (\ref{eq:dic}).\par

Let us turn to the three--quark system. In this
 case  the basic quantity is
\begin{eqnarray}
  W_{3q} = \frac{1}{3!} \left\langle \varepsilon_{a_1 a_2 a_3}
 \varepsilon_{b_1 b_2 b_3}  \left[ {\rm  P} \exp \left( ig
 \int_{\overline{\Gamma}_1} dx^{\mu_1} A_{\mu_1}(x) \right)
\right]^{a_1 b_1} \right.
\nonumber\\
\left. \left[ {\rm P} \exp \left( ig \int_{\overline{\Gamma}_2}
 dx^{\mu_2} A_{\mu_2}(x) \right) \right]^{a_2 b_2}
  \left[  {\rm P} \exp \left( ig \int_{\overline{\Gamma}_3 } dx^{\mu_3}
A_{\mu_3}(x) \right) \right]^{a_3 b_3} \right\rangle \> .
\label{eq:w3}
\end{eqnarray}
\noindent Here $a_j,b_j$ are colour indices, $j=1,2,3$ and
$\overline{\Gamma}_j$  denote the  curve made by:
the world lines $\Gamma_j$ for the  quark $j$
between the times $t_{\rm i}$ and $t_{\rm f}$
($t_{\rm i} < t_{\rm f}$), a straight  line
on the surface $t=t_{\rm i}$ merging from an arbitrary
fixed  point $I$ (which
 we also denote by $y_M$) and connected to
the world line, another  straight line
on the surface $t=t_{\rm f}$ connecting
the world line to a second fixed point $F$ (also denoted  as $x_M$).
\par
Under the  assumptions i) and ii) we can write
in place of (\ref{eq:sex}) and ({\ref{eq:dodici})
\begin{equation}
 i \ln W_{3q} = \frac{2}{3} g^2 \sum _{i<j}
 \int _{\Gamma _i} dx^{\mu}
_i \int _{\Gamma _j} dx^{\nu}_j \  i D_{\mu \nu} (x_i - x_j)
+ \sigma S_{\min} + {1\over 3} C P \> .
\label{eq:diciannove}
\end{equation}
Here the perturbative term is taken at the lowest order in $\alpha_s$
 and $S_{\min}$ denotes the minimum among all
 the surfaces made by three sheets
having the curves $\overline{\Gamma}_1$,
 $\overline{\Gamma}_2$ and $\overline{\Gamma}_3$ as contours and
joining on a line $\Gamma_M$ connecting $I$ with $F$ (the minimum
 is understood at fixed $
\bar{\Gamma}_j$
 as the surfaces and $\Gamma_{\rm M}$ change ).
 Obviously, $P$ denotes the total length of $\bar{\Gamma}_1$,
 $\bar{\Gamma}_2$ and $\bar{\Gamma}_3$.
Notice that a priori the constants $\sigma$ and  $C$
occurring in (17)
 could be  different from those occurring in (\ref{eq:dodici});
 however, the fact that when two quarks coincide
 the potential
 derived from (\ref{eq:diciannove}) must coincide with that derived from
 (\ref{eq:dodici})  ( in a colour singlet
 state two quarks are equivalent to an antiquark) grants that they must be
actually equal.\par
The right--hand  side of (\ref{eq:diciannove})  can be evaluated
 as an expansion  in $\dot{\bf z}_j$  on the same foot used for
 Eqs.(\ref{eq:otto}), (\ref{eq:undici}) and (\ref{eq:dic}). In particular up to
 the second order in the velocities, $ S_{\rm min}$ coincides with the
 surface described by the equations
\begin{equation}
{\bf x}_j^{\rm min} (t,s) =s {\bf z}_j (t) + (1-s) {\bf z}_{M}(t)\> ,
\      \    \    \qquad \qquad  j=1,2,3 \> .
\end{equation}
Here $ {\bf z}_{M} (t)$  is
 constructed from the positions ${\bf z}_1 (t)$, $ {\bf z}_2 (t)$ and
 ${\bf z}_3 (t)$ of the three quarks according to the following rule:
 if no angle in the
triangle   made by ${\bf z}_1 (t)$, $ {\bf z}_2 (t)$ and
 ${\bf z}_3 (t)$ exceeds $120^0$ (configuration I), ${\bf z}_M (t)$
 coincides with the point inside the triangle which sees  the three sides
 under the same angle $120^0$; if one of the three angles in the
triangle  is $\ge 120^0$ (configuration II), ${\bf z}_{M} (t)$
 coincides with the corresponding vertex, let us say ${\bf z}_{\bar{j}}(t)$.
\par
In conclusion, the result is
\begin{eqnarray}
i \ln W_{3q} = \int_{t_{\rm i}}^{t_{\rm f}} dt \   \left\{ \sum_{j<l}
\left[ - \frac{2}{3} \frac{\alpha_s}{r_{jl}} + \frac{1}{2} \frac{2}{3}
\frac{\alpha_s}{r_{jl}} (\delta^{hk} + \hat{r}_{jl}^h \hat{r}_{jl}^k)
\dot{z}_j^h \dot{z}_l^k \right] + \right.
\nonumber\\
\left. {}+ \sigma \sum_{j=1}^{3} r_j \left[ 1-\frac{1}{6}
 (\dot{{\bf z}}_{j{\rm T}_j}^2 + \dot{{\bf z}}^2_{M {\rm T}_j}
+ \dot{{\bf z}}_{j{\rm T}_j} \cdot
\dot{{\bf z}}_{M {\rm T}_j}) \right]
+{C\over 3} \sum_{j=1}^{3} \int_{t_{\rm i}}^{t_{\rm f}}
dt (1 -{1\over 2} \dot{z}_j^h \dot{z}_j^h)
 \right\},
\label{eq:ventidue}
\end{eqnarray}
where  ${\bf r}_{jl} = {\bf r}_j -{\bf r}_l \equiv {\bf z}_j - {\bf
z}_l$, $ {\bf r}_j ={\bf z}_j -{\bf z}_{M}$
 and the transversal prescription ${\rm T}_j$ is now referred
 to ${\bf r}_j$. Furthermore we can notice that
the quantity $\dot{\bf z}_M$ can be obtained by deriving
 the equation $\sum_{j=1}^3 ({\bf r}_j / r_j)=0$. We have
 indeed
 $\sum_{j=1}^3 {1\over r_j} (\delta^{hk} -\hat{r}_j^h
\hat{r}_j^k) \dot{z}_j^k =\sum_{j=1}^3 {1\over r_j} (\delta^{hk} -\hat{r}_j^h
\hat{r}_j^k) \dot{z}_M^k$. Obviously in  configuration II  we
have $\dot{\bf z}_M =\dot{\bf z}_{\bar{j}}$.

\section{Quark--antiquark potential}

The starting point is the gauge invariant quark-antiquark
$(q_1,\bar{q}_2)$ Green function
(for definiteness let us  assume the two  particles
 to have
 different flavours)
\begin{eqnarray}
& & G(x_1,x_2;y_1,y_2) =
\frac{1}{3}\langle0|{\rm T}\psi_2^c(x_2)U(x_2,x_1)\psi_1(x_1)
\overline{\psi}_1(y_1)U(y_1,y_2)  \overline{\psi}_2^c(y_2)
|0\rangle =
\nonumber\\
& &= \frac{1}{3} {\rm Tr} \langle U(x_2,x_1)
 S_1^{{\rm F}}(x_1,y_1|A) U(y_1,y_2) C^{-1}
S_2^{{\rm F}}(y_2,x_2|A) C \rangle \> ,
\label{eq:ventitre}
\end{eqnarray}
where $c$ denotes the charge-conjugate fields, $C$ is the charge-conjugation
matrix, $U$
the path-ordered gauge string
$U(b,a)= {\rm P}  \exp  \left(ig\int_a^b dx^{\mu} \, A_{\mu}(x) \right)$
(the integration path being the straight line joining $a$ to $b$),
 $S_1^{{\rm F}}$ and $S_2^{{\rm F}}$ the quark propagators in an
external gauge field $A^{\mu}$.

We assume $x_1^0=x_2^0=t_{\rm f}$,
$y_1^0=y_2^0=t_{\rm i}$ (with $t_{\rm f}-t_{\rm i} >0$
and large) and
note that $S_j^{{\rm F}}$ are $4\times4$  Dirac type matrices.
Then, performing a Foldy--Wouthuysen transformation on $G$,
we can replace $S_j^{{\rm F}}$ with a Pauli propagator $K_j$ (a
$2\times2$ matrix in the spin indices) and obtain a two-particle
Pauli-type
Green function $K$.
Solving the
Schr\"odinger-like equation for $ K_j$ by the path--integral technique and
 replacing it in the expression of $K$,
 we obtain  even this quantity
 in the form
of a path integral on the world lines of the two quarks
 (see Ref.$^1$ for details):
\begin{eqnarray}
& & K({\bf x}_1, {\bf x}_2, {\bf y}_1, {\bf y}_2; t_{\rm f} - t_{\rm i})=
\int_{{\bf z}_1(t_{\rm i})={\bf y}_1}^{{\bf z}_1(t_{\rm f})={\bf x}_1}
    {\cal D}  [{\bf z}_1, {\bf p}_1]
\int_{{\bf z}_2(t_{\rm i})={\bf y}_2}^{{\bf z}_2(t_{\rm f})={\bf x}_2}
    {\cal D}   [{\bf z}_2, {\bf p}_2]\nonumber \\
& &  \exp \{i\int_{t_{\rm i}}^{t_{\rm f}} dt\,
 \sum_{j=1}^2
[{\bf p}_j \cdot \dot{{\bf z}}_j -m_j-
\frac{{\bf p}^2_j}{2m_j}+\frac{{\bf p}^4_j}{8m_j^3} ]\}
\langle \frac{1}{3}
{\rm Tr \, T_s \, P} \exp \{ig\oint_{\Gamma} dx^{\mu} \,
A_{\mu}(x) \nonumber \\
& & +\sum_{j=1}^2\frac{ig}{m_j} \int_{{\Gamma}_j} dx^{\mu}
  (S_j^l \hat{F}_{l{\mu}}(x) -\frac{1}{2m_j}
S_j^l\varepsilon^{lkr}p_j^k F_{{\mu}r}(x)-
\frac{1}{8m_j} D^{\nu}F_{{\nu}{\mu}}(x)
 ) \}  \rangle \> .
\label{eq:path1}
\end{eqnarray}
Here ${\rm T_s}$ is the time-ordering prescription for the
 spin matrices; P,
 ${\rm Tr}$, $\Gamma$, $\Gamma_1$ and $ \Gamma_2$  are defined as in Eq.(1).
 Furthermore, as usual
$F^{\mu\nu}= \partial^{\mu} A^{\nu} - \partial^{\nu} A^{\mu}
+ ig[A^{\mu},A^{\nu}]$,
$\hat{F}^{\mu\nu}= \frac{1}{2} \varepsilon^{\mu\nu\rho\sigma}
F_{\rho\sigma}$ and
$D^{\nu}F_{\nu\mu}= \partial^{\nu}F_{\nu\mu}+
ig[A^{\nu},F_{\nu\mu}]\>$,
 $\varepsilon^{\mu\nu\rho\sigma}$ being the four-dimensional Ricci
symbol. \par
In order to show that the interaction between $q_1$ and $\bar{q}_2$
 can be described in terms of a semirelativistic potential
 we must check that at the order $ {1\over m^2}$ we can write
\begin{equation}
\left\langle \frac{1}{3}
{\rm Tr \, T_s \, P} \exp \ldots\right \rangle =
{\rm T_s} \exp\left[ -i\int_{t_{\rm i}}^{t_{\rm f}} dt \,
 V^{q\overline{q}}({\bf z}_1(t),{\bf
z}_2(t),{\bf p}_1(t),{\bf p}_2(t),{\bf S}_1,{\bf S}_2) \right] \> ,
\label{eq:ventinove}
\end{equation}
for some
$V^{q\overline{q}}$.
Expanding the logarithm on the left-hand side of (\ref{eq:ventinove}),
this is equivalent to state
\begin{eqnarray}
 i \ln W_{q\overline{q}}  +
i \sum_{j=1}^2 \frac{ig}{m_j} \int_{{\Gamma}_j}dx^{\mu}
\bigg( S_j^l \, \langle\langle \hat{F}_{l\mu}(x)
 \rangle \rangle -\frac{1}{2m_j} S_j^l
 \varepsilon^{lkr} p_j^k \, \langle\langle
F_{\mu r}(x) \rangle\rangle -
\nonumber\\
{} - \frac{1}{8m_j} \, \langle\langle
D^{\nu} F_{\nu\mu}(x) \rangle\rangle \bigg)
 - \frac{1}{2} \sum_{j,j^{\prime}} \frac{ig^2}{m_jm_{j^{\prime}}}
{\rm T_s} \int_{{\Gamma}_j} dx^{\mu} \, \int_{{\Gamma}_{j^{\prime}}}
dx^{\prime\sigma}
\, S_j^l \, S_{j^{\prime}}^k
\nonumber\\
\bigg( \, \langle\langle \hat{F}_{l \mu}(x)
 \hat{F}_{k \sigma}(x^{\prime})
\rangle\rangle - \, \langle\langle \hat{F}_{l \mu}(x) \rangle\rangle
\, \langle\langle \hat{F}_{k \sigma}(x^{\prime}) \rangle\rangle \bigg)
 +\dots =\left[\int_{t_{\rm i}}^{t_{\rm f}} dt
 \, V^{q\overline{q}} \right] \> ,
\end{eqnarray}
with the notation
\begin{equation}
\langle\langle f[A] \rangle\rangle = \frac{\frac{1}{3} \left\langle
{\rm Tr \, P}\left[ \exp ( ig\oint_{\Gamma}dx^{\mu}
 \, A_{\mu}(x) ) \right] f[A]
\right\rangle}
{\frac{1}{3} \left\langle {\rm Tr \, P}  \exp ( ig\oint_{\Gamma} dx^{\mu}
\, A_{\mu}(x) ) \right\rangle}
\end{equation}
and $W_{q\overline{q}}$ as given by (\ref{eq:dicbis}).\par
 Notice that after replacing $\dot{\bf z}_j$ by $ {{\bf p}_j \over m_j}$
 in Eqs.(\ref{eq:otto}), (\ref{eq:undici}) and (\ref{eq:dic}),
 the expression resulting for
  $ i \ln W_{ q \bar{q}}$
 is already of the desired form. Concerning the  spin--dependent part
 we observe that the occurring field expectation values  can
 be expressed in terms of $ i \ln W_{q \bar{q}}$ by the functional derivatives
\begin{equation}
g \, \langle \langle F_{\mu\nu}(z_1) \rangle \rangle =
\frac{ \delta (i \ln W_{q\overline{q}}) }
{ \delta S^{\mu\nu}(z_1)} \>,
\label{eq:trentatre}
\end{equation}

\begin{eqnarray}
g^2 \, \bigg( \langle \langle F_{\mu\nu}(z_1)
F_{\rho\sigma}(z_2) \rangle \rangle
& -& \langle \langle F_{\mu\nu}(z_1) \rangle \rangle \,
\langle \langle F_{\rho\sigma}(z_2)  \rangle \rangle \bigg) =
\nonumber\\
& = &\frac{ \delta^2 \ln W_{q\overline{q}}}
{\delta S^{\mu\nu}(z_1) \delta S^{\rho\sigma}(z_2)}
 =  - ig \frac{\delta}{\delta S^{\rho\sigma}(z_2)}
\, \langle \langle F_{\mu\nu}(z_1) \rangle \rangle ,
\label{eq:trentaq}
\end{eqnarray}
where $ \delta S^{\mu \nu} (z_j) ={1\over 2} ( d z_j^{\mu} \delta z_j^{\nu}
 - d z_j^{\nu} \delta z_j^{\mu})$ is the element of the  surface  spanned
 by the path $ z_j(t)$ as a consequence of the variation
 $ z_j (t) \to z_j (t) +\delta z_j (t)$.\par
The evaluation of the right hand side of (\ref{eq:trentatre})
 and (\ref{eq:trentaq}) requires some care, since the functional
 derivatives may lower the order of magnitude in the velocities.
 However, it
 can be done  without any additional
 assumptions and the results are
\begin{equation}
g \, \langle\langle F_{0k}(z_1) \rangle\rangle =
( \frac{4}{3} \alpha_s \frac{1}{r^3}+ \frac{\sigma}{r}) r^k
+ O(v^2)\> ,
\end{equation}
\begin{equation}
g \, \langle\langle F_{hk}(z_1) \rangle\rangle =
 (\frac{4}{3} \frac{\alpha_s}{m_2} \frac{1}{r^3}
+ \frac{\sigma}{m_2} \frac{1}{r} )
( r^h p_2^k - r^k p_2^h ) +O(v^3) \> ,
\end{equation}
\begin{eqnarray}
& & g^2 \, \bigg( \langle\langle F_{hk}(z_1) F_{lm}(z_2) \rangle\rangle -
\langle\langle F_{hk}(z_1) \rangle\rangle \,
\langle\langle F_{lm}(z_2) \rangle\rangle \bigg)  =
\nonumber\\
& &= - \frac{4}{3} \frac{ig^2}{8\pi} \delta (t_1-t_2) \left\{
\partial_l \partial_k \left[ \frac{1}{r} \left( \delta^{hm}
+ \hat{r}^h \hat{r}^m \right)\right] - \partial_l \partial_h \left[
\frac{1}{r} \left( \delta^{km} + \hat{r}^k
 \hat{r}^m \right) \right]
- \right.
\nonumber\\
& &\left. - \partial_m \partial_k \left[ \frac{1}{r} \left(
\delta^{hl} + \hat{r}^h \hat{r}^l \right)\right] + \partial_m
\partial_h \left[ \frac{1}{r} \left( \delta^{kl} + \hat{r}^k
\hat{r}^l \right)\right] \right\} + O(v^2)\> ,
\end{eqnarray}

\begin{equation}
\bigg( \langle\langle F_{\mu\nu}(z_1) F_{\rho\sigma}(z_1')
\rangle\rangle - \langle\langle F_{\mu\nu}(z_1) \rangle\rangle \,
\langle\langle F_{\rho\sigma}(z_1') \rangle\rangle \bigg)
=0
\end{equation}
and similar ones. \par
In the end one obtains the potential in the form  of a static part,
a spin--dependent part  and a velocity--dependent one,
$ V^{q\overline{q}} = V^{q\overline{q}}_{{\rm stat}} +
V^{q\overline{q}}_{{\rm sd}} +
V^{q\overline{q}}_{{\rm vd}}$,
with
\begin{equation}
V^{q\overline{q}}_{{\rm stat}} = -  \frac{4}{3}
 \frac{{\alpha}_s}{r} + \sigma r -{4\over 3} {\alpha_{\rm s}^2 \over 4 \pi}
 {1\over r} [{66- 4 N_{f} \over 3}
 (\ln \mu r + \gamma) +A]
\label{eq:trena}
\end{equation}
\begin{eqnarray}
& & V^{q\overline{q}}_{{\rm sd}} =
\frac{1}{8} \sum_{j=1,2} {1\over m_j^2}
\nabla^2 \left( - \frac{4}{3} \frac{\alpha_s}{r} + \sigma r \right)
+
   \left(
 \frac{4}{6} \frac{\alpha_s}{r^3} -
\frac{\sigma}{2 r} \right) \sum_{j=1,2} \frac{1}{m_j^2} {\bf S}_j \cdot
{\bf L}_j
 + \frac{1}{m_1m_2} \frac{4}{3} \frac{\alpha_s}{r^3}\nonumber \\
& & ({\bf S}_2
\cdot{\bf L}_1  + {\bf S}_1 \cdot{\bf L}_2 ) +
 {4 \alpha_{\rm s}\over 3 m_1 m_2} \left [
 (\frac{3}{r^5} ({\bf S}_1 \cdot {\bf r})({\bf S}_2 \cdot
{\bf r}) - {{\bf S}_1 \cdot {\bf S}_2\over r^3}) +
\frac{8\pi}{3} \delta^3({\bf r}) {\bf S}_1 \cdot {\bf S}_2 \right ]
\label{eq:trenb}
\end{eqnarray}

\begin{eqnarray}
V_{\rm vd}^{q \bar{q}} & =& {1\over m_1 m_2} \{ {4\over 3}
 {\alpha_s \over r} ( \delta^{hk}+ \hat{r}^h \hat{r}^k)
 p_1^h p_2^k \}_{\rm Weyl} - C \sum_j {p_j^2 \over 4 m_j^2} \nonumber \\
& -& \sum_{j=1}^2 {1\over 6 m_j^2} \{ \sigma r {\bf p}_{j {\rm T}}
\}_{\rm Weyl} -{1\over 6 m_1 m_2} \{ \sigma r {\bf p}_{1 {\rm T}}
 \cdot {\bf p}_{2 {\rm T}} \}_{\rm Weyl}.
\label{eq:trenc}
\end{eqnarray}

At  the order $ \alpha_s$ the above  potential
 coincides globally  with that given in Ref.$^1$. In particular
$V_{\rm sd}$
 was  originally given by Eichten and Feinberg  and corrected by
 Gromes$^2$ (see also Ref.$^3$), while $V^{\rm vd}$
  has been obtained for the first time in$^1$.
Notice  that Eq.(\ref{eq:trenc}) differs from the
 corresponding one
   proposed under the ad hoc assumption
 of scalar confinement and does not present the phenomenological
 difficulties of this $^7$. Notice also that the terms in $C$ can
 be reabsorbed in a redefinition of the masses $m_j\to m_j'=
m_j+{C\over 2}$.
 The $O(\alpha_s^2)$ term in
 (\ref{eq:trena}) has been obtained for the first time in $^6$.\par
The  $ O(\alpha_s^2)$ contributions to $V_{\rm sd}$
 and $V_{\rm vd}$  have been evaluated by Gupta et al.$^8$
 in an $S$ matrix  context  but they  have not been
 included here.
 In fact  such contributions
 are found to be important for an understanding of the fine and hyperfine
 structure  of the meson spectrum. However, due to the ambiguities inherent
  in the derivation method, a consistent evaluation in the Wilson loop
 approach should be desirable. Calculations are in progress
 in this line.\par
Finally, let us come to the ordering   in (\ref{eq:trenc}).
 Obviously, ordering is related to the discretization prescription
 in the definition of the path integral. If in the definition
of the gauge field functional integration we identify  the element
 $U_{n' n}$ of the colour group associated to the link between
 the contiguous sites $n$ and $n'$ with $\exp{ [ig (x_{n'}-x_{n})^{\mu}
 A_{\mu} ({x_{n'}+ x_{n} \over 2})]}$, we obtain the Weyl ordering
\begin{eqnarray}
& & \quad \quad \{ X^{hk}(r), p^h_j p^k_{j\prime} \}_{\rm Weyl} = {1\over 4}
 \{p^h_j, \{ X^{hk}(r), p^k_{j\prime}\} \}=\nonumber \\
& & = {1\over 4} ( X^{hk} (r) p^h_j p^k_{j\prime}+
   p^h_j X^{hk}(r) p^k_{j\prime} +  p^k_{j\prime} X^{hk} (r) p^h_j +
 X^{hk} (r) p^h_j p^k_{j\prime}).
\label{eq:ordering}
\end{eqnarray}

\section{Three-quark potential}

The three-quark
 gauge invariant Green function  can be written as
(again we assume the quarks to have differents flavours)
\begin{eqnarray}
& & \qquad   \qquad \quad \quad G(x_1,x_2,x_3,y_1,y_2,y_3)
 = \frac{1}{3!} \varepsilon_{a_1 a_2 a_3}
\varepsilon_{b_1 b_2 b_3}
\nonumber\\
& &\langle 0| {\rm T} \,
U^{a_3 c_3}(x_M,x_3) U^{a_2 c_2}(x_M,x_2)
 U^{a_1 c_1}(x_M,x_1)
\psi_{3 c_3}(x_3)
\psi_{2 c_2}(x_2) \psi_{1 c_1}(x_1)
\nonumber\\
& &\overline{\psi}_{1 d_1}(y_1) \overline{\psi}_{2 d_2}(y_2)
\overline{\psi}_{3 d_3}(y_3)
U^{d_1 b_1}(y_1,y_M) U^{d_2 b_2}(y_2,y_M)
 U^{d_3 b_3}(y_3,y_M)
 |0 \rangle
\label{eq:gtre}
\end{eqnarray}
and we assume
 $x_1^0=x_2^0=x_3^0= x_M^0= t_{\rm f}$,
$y_1^0=y_2^0=y_3^0= y_M^0=t_{\rm i}$,
$  t_{\rm f}- t_{\rm i}$ large.

The integration over the fermionic variables is again trivial
 and one can write
\begin{eqnarray}
G({\bf x}_1, {\bf x}_2, {\bf x}_3, {\bf y}_1, {\bf y}_2,
{\bf y}_3; \tau)
 = \frac{1}{3!} \varepsilon_{a_1 a_2 a_3}
\varepsilon_{b_1 b_2 b_3} \bigg\langle \bigg( U(x_M,x_1)
S_1^{\rm F}(x_1,y_1|A) U(y_1,y_M) \bigg)^{a_1 b_1}
\nonumber\\
\bigg( U(x_M,x_2) S_2^{\rm F}(x_2,y_2|A)
 U(y_2,y_M) \bigg)^{a_2 b_2}
\bigg( U(x_M,x_3) S_3^{\rm F}(x_3,y_3|A)
 U(y_3,y_M) \bigg)^{a_3 b_3}
\bigg\rangle \> .
\label{eq:treprop}
\end{eqnarray}

{}From (\ref{eq:treprop}),
we can proceed  strictly as in Sec.3 and
in conclusion we have to show that
\begin{eqnarray}
 i \ln W_{3q}  +
i \sum_{j=1}^3 \frac{ig}{m_j} \int_{{\Gamma}_j}dx^{\mu}
\bigg( S_j^l \, \langle\langle \hat{F}_{l\mu}(x)
 \rangle \rangle -\frac{1}{2m_j} S_j^l
 \varepsilon^{lkr} p_j^k \, \langle\langle
F_{\mu r}(x) \rangle\rangle -
\nonumber\\
{} - \frac{1}{8m_j} \, \langle\langle
D^{\nu} F_{\nu\mu}(x) \rangle\rangle \bigg)
 - \frac{1}{2} \sum_{j,j^{\prime}} \frac{ig^2}{m_jm_{j^{\prime}}}
{\rm T_s} \int_{{\Gamma}_j} dx^{\mu}
\, \int_{{\Gamma}_{j^{\prime}}}
dx^{\prime\sigma}
\, S_j^l \, S_{j^{\prime}}^k \cdot
\nonumber\\
\cdot \bigg( \, \langle\langle \hat{F}_{l \mu}(x)
 \hat{F}_{k \sigma}(x^{\prime})
\rangle\rangle - \, \langle\langle \hat{F}_{l \mu}(x) \rangle\rangle
\, \langle\langle \hat{F}_{k \sigma}(x^{\prime}) \rangle\rangle \bigg)
 =  \left[\int_{t_{\rm i}}^{t_{\rm f}} dt
 \, V^{3q}({\bf z}_j, {\bf p}_j, {\bf S}_j) \right]
\end{eqnarray}
with  $W_{3q}$ given by (\ref{eq:ventidue}) and
\begin{equation}
\langle\langle f[A] \rangle\rangle= \frac{ \frac{1}{3!} \langle
\varepsilon \varepsilon \, \{ \prod_j \, {\rm P}
[ \exp (ig \int_{\overline{\Gamma}_j} dx^{\mu}
\, A_{\mu}(x) )] \} f[A] \rangle }{\frac{1}{3!} \langle \varepsilon
\varepsilon \, \{ \prod_j \, {\rm P} \exp
(ig \int_{\overline{\Gamma}_j} dx^{\mu} \, A_{\mu}(x)
) \} \rangle } \> .
\end{equation}
Again, after the replacement $ \dot{{\bf z}}_j \to {\bf p}_j/m_j$,
the quantity  $ i \ln W_{3 q}$  is
 already of the desired form, while  the field expectation values
 can be evaluated according to equations analogues to (\ref{eq:trentatre})
 and (\ref{eq:trentaq}) and lead to similar expressions.
The final result is again of the form
$ V^{3q} = V^{3q}_{{\rm stat}} + V^{3q}_{{\rm sd}} +
V^{3q}_{{\rm vd}}$
with
\begin{eqnarray}
V^{3q}_{{\rm stat}} & =&
\sum_{j<l} \left( -\frac{2}{3}
\frac{\alpha_s}{r_{jl}} \right) + \sigma (r_1+r_2+r_3)+
C,
\label{eq:trestat}\\
V^{3q}_{{\rm sd}} & = &
\frac{1}{8m_1^2} \nabla^2_{(1)} \left( -\frac{2}{3}
\frac{\alpha_s}{r_{12}} -\frac{2}{3} \frac{\alpha_s}{r_{31}} +
\sigma r_1 \right) +
\nonumber\\
{}&+& \left\{ \frac{1}{2m_1^2} {\bf S}_1 \cdot \left[
({\bf r}_{12} \times {\bf p}_1) \left( \frac{2}{3}
\frac{\alpha_s}{r_{12}^3} \right) + ({\bf r}_{31} \times
{\bf p}_1) \left( -\frac{2}{3} \frac{\alpha_s}{r_{31}^3}
\right) - \frac{\sigma}{r_1} ({\bf r}_1 \times {\bf p}_1)
\right]  \right. +
\nonumber\\
{} &+& \left. \frac{1}{m_1m_2} {\bf S}_1 \cdot ({\bf r}_{12}
\times {\bf p}_2) \left( -\frac{2}{3} \frac{\alpha_s}{r_{12}^3}
\right) + \frac{1}{m_1m_3} {\bf S}_1 \cdot ({\bf r}_{31}
\times {\bf p}_3) \left( \frac{2}{3} \frac{\alpha_s}{r_{31}^3}
\right) \right\} +
\nonumber\\
{} &+&
 \frac{1}{m_1m_2} \frac{2}{3} \alpha_s \left\{
\frac{1}{r_{12}^3} \left[ \frac{3}{r_{12}^2} ({\bf S}_1
\cdot {\bf r}_{12}) ({\bf S}_2 \cdot {\bf r}_{12}) -
{\bf S}_1 \cdot {\bf S}_2 \right] + \frac{8\pi}{3}
\delta^3 ({\bf r}_{12}) {\bf S}_1 \cdot {\bf S}_2
\right\}  +
\nonumber\\
{} &+& \, \mbox{cyclic permutations} \> ,
\label{eq:tresd}
\end{eqnarray}
\begin{eqnarray}
V^{3q}_{{\rm vd}} &=&
 \sum_{j<l} \frac{1}{2m_jm_l} \left\{ \frac{2}{3}
\frac{\alpha_s}{r_{jl}} (\delta^{hk} + \hat{r}_{jl}^h \hat{r}_{jl}^k)
p_j^h p_l^k \right\}_{{\rm Weyl}} -
\sum_{j=1}^{3} \frac{1}{6m_j^2} \{ \sigma \, r_j \,
 {\bf p}_{j{\rm T}_j}^2 \}_{{\rm Weyl}} -
\nonumber\\
{} &-& \sum_{j=1}^{3} \frac{1}{6} \{ \sigma \, r_j \,
 \dot{{\bf z}}_{M{\rm T}_j}^2 \}_{{\rm Weyl}} -
\sum_{j=1}^{3} \frac{1}{6m_j} \{ \sigma \, r_j \,
 {\bf p}_{j{\rm T}_j} \cdot \dot{{\bf z}}_{M{\rm T}_j} \}_{{\rm Weyl}}
- \sum_{j} {C \over  6 m_j^2} p^2_j,
\>
\label{eq:trevd}
\end{eqnarray}
where the notations are the same as used in (\ref{eq:ventidue})
 and the ordering is as in (\ref{eq:ordering}).
Notice, in particular, that
 the quantity $\dot{{\bf
z}}_{M}$
 in (\ref{eq:trevd}) is given by
\begin{equation}
 \dot{\bf z}_{M} = \left\{
\begin{array} {ll}
R^{-1} \sum_{j=1}^3 \left( {\bf p}_{j {\rm T}_j}/m_j r_j   \right)
\qquad \mbox{ type I configuration}& \\
{\bf p}_{\bar{j}}/m_{\bar{j}} \qquad \qquad
\qquad \qquad \quad \mbox{ type II configuration}:
({\bf z}_M\equiv{\bf z}_{\bar{j}}) \> ,
\end{array} \right.
\end{equation}
$R$ being the matrix with elements $R^{hk}= \sum_{j=1}^3 \frac{1}{r_j}
(\delta^{hk} -\hat{r}^h_j \hat{r}^k_j)$.

Notice  that Eq.(\ref{eq:tresd}) properly refers to   the configuration
I case.
 In general one should write $
V_{\rm sd}^{\rm LR}= - \sum_{j=1}^3 {1 \over 2 m_j^2}
 {\bf S}_j \cdot {\bf \nabla}_{j} V_{\rm stat}^{\rm LR} \times
 {\bf p}_j $
(In comparing this with (\ref{eq:tresd})
 one should keep in mind that the
partial derivatives in ${\bf z}_M$ of
$V_{\rm stat}^{\rm LR}$ vanish due to the definition of $M$).

We observe that
the short range part in Eqs.(\ref{eq:trestat})--(\ref{eq:trevd})
 is of  a  pure two body
type: in fact it is identical to the electromagnetic
 potential among three equally charged
particles but for the colour group factor $2/3$ and
it is well known. Even the static
confining potential in Eq.(\ref{eq:trestat}) is well  known
(for a review see e.g.$^9$).
Furthermore the long range part  in
Eq.(\ref{eq:tresd}) coincides with the expression obtained by Ford$^{10}$
 starting
from the assumption of a purely scalar Salpeter potential of the form
\begin{equation}
 \sigma \, (r_1 + r_2 + r_3) \, {\beta}_1 {\beta}_2 {\beta}_3 \> ,
\label{eq:fordbs}
\end{equation}
but to our knowledge it was not  obtained consistently in a
Wilson loop context before Ref.$^1$. Eq.(\ref{eq:trevd}) has
been given for the first time in Ref.$^1$. Eq. (\ref{eq:trevd})
differs from  the corresponding equation obtained from (\ref{eq:fordbs}).
The situation for  the three quarks is so similar to that occurring for
 the quark--antiquark system. In Eqs.(39) and (41) the terms in $C$
 can be again eliminated by  the redefinition of the masses
 $m_j\to m_j''=m_j+{C\over 3}$. Notice however  that $m_j''$ differs
 from $m_j'$.

\section{ Relativistic flux tube model}

Let us now neglect in Eq.(\ref{eq:path1}) the spin--dependent terms
 and replace the ${1\over m^2}$ expansion by its exact relativistic
 expression
\begin{eqnarray}
& &K({\bf x}_1, {\bf x}_2;  {\bf y}_1, {\bf y}_2; t_{\rm f} - t_{\rm i}) =
\nonumber \\
& &\int {\cal D} [{\bf z}_1, {\bf p}_1]
\int {\cal D} [{\bf z}_2, {\bf p}_2]
\exp \left\{i \left[ \int_{t_{\rm i}}^{t_{\rm f}} dt \sum_{j=1}^2 ({\bf p}_j
\cdot {\dot{\bf z}_j} - \sqrt{m_j^2 + {\bf p}_j^2})
\right] + \ln W_{q \bar{q}} \right\} .
\label{eq:alfa}
\end{eqnarray}
Let us further evaluate $ i \ln W_{ q \bar{q}}^{\rm SR}$ by the original
   Eq.(\ref{eq:sex}) and assume that a sensible
 approximation is obtained even in the  relativistic case postulating
  the first line  of Eq.(\ref{eq:dic}) in the center--of--mass system
 of the two particles. Then, if we expand again the exponent in
 (\ref{eq:alfa}) around the stationary values ${\bf p}_j = { m {\bf z}_j
\over \sqrt{1-{\bf z}^2_j}}$, in the gaussian approximation
we obtain the ordinary  lagrangian
\begin{eqnarray}
 L  & & = - \sum_{j=1}^2 m_j \sqrt{1- \dot {\bf z}_j^2 } +
 {4 \over 3} { {\alpha}_s  \over r} \left[ 1 - \frac{1}{2}
( \delta^{hk} + \hat{r}^h \hat{r}^k ) \dot{z}_1^h \dot{z}_2^k \right]
+\nonumber \\
 & & {} - \sigma
r \int_0^1 ds [1 - ( s \dot{\bf z}_{1{\rm T}} + (1-s) \dot{{\bf
z}}_{2{\rm T}} )^2 ]^{1/2}.
\label{eq:beta}
\end{eqnarray}
This coincides with the relativistic flux--tube lagrangian$^{11}$.\par
 From (\ref{eq:beta}) is not possible to obtain even a classical
 hamiltonian in a closed form, due to the complicate velocity dependence.
However, in terms of an expansion in ${\sigma \over m^2}$ we have  ( we
 assume $m_1=m_2=m$ for simplicity  and have already  eliminated
 the terms in $C$)
\begin{eqnarray}
{\cal{H}}({\bf r}, {\bf q})& = &2 \sqrt{m^2+q^2} +{\sigma r\over 2}
 \Big [{\sqrt{m^2+q^2}\over q_{\rm T}}\, {\rm arcsin}{q_{\rm T}
 \over \sqrt{m^2+q^2}}
+\sqrt{{m^2+ q_{\rm r}^2\over m^2+q^2}}\Big ]+\nonumber \\
& +&{\sigma^2 r^2 \over 16 q_{\rm T}^2} {m^2+ q_{\rm r}^2 \over \sqrt{m^2+q^2}}
\Big  [ {\sqrt{m^2+q^2} \over q_{\rm T}}\,  {\rm arcsin} {q_{\rm T}
 \over \sqrt{m^2+q^2}}
-\sqrt{{m^2+q_{\rm r}^2\over m^2+q^2}}\Big ]^2 + \dots
\label{eq:gamma}
\end{eqnarray}
with ${\bf r}= {\bf z}_{1{\rm CM}}- {\bf z}_{2{\rm CM}}$,
${\bf q}={\bf p}_{1 {\rm CM}}=-{\bf p}_{2 {\rm CM}}$, ${\bf q}_{\rm r}=
(\hat{\bf r}\cdot {\bf q})/ {\hat{\bf r}}$ and  $q^h_{\rm T}= (\delta^{hk}
 -\hat{r}^h \hat{r}^k ) q^k$.
{}From this a quantum hamiltonian can  be immediately obtained
 by setting
\begin{equation}
 \langle  {\bf k}' \vert H_{\rm FT} \vert {\bf k}\rangle=
 \int {{\rm d}{\bf r} \over (2 \pi)^3} e^{i ({\bf k}-{\bf k}')
\cdot {\bf r}} \,\,
 {\cal {H}}({\bf r}, {{\bf k}'+
{\bf k}\over 2}),
\label{eq:delta}
\end{equation}
in which the ordering prescription is again Weyl prescription.
 By an expansion in ${1\over m^2}$ a semirelativistic
 hamiltonian  can be obviously reobtained with a potential
 given by (\ref{eq:trena})--(\ref{eq:trenc}).

\section{ Bethe--Salpeter equation}

Let us go back  to the
equation analogous to (\ref{eq:ventitre})
 for spinless quarks
 and  in it use  the covariant representation for the quark propagator
 in an external gauge field
\begin{equation}
\Delta^{\rm F} (x,y\vert A)={- i\over 2} \int_0^{\infty} d \tau
 \int_{z(0)=y}^{z(\tau)=x} {\cal D} [z] {\rm P}
{\rm exp}\,  i\int_0^{\tau} {\rm d} \tau\prime \{ -{1\over 2}
 [ ({d z \over d \tau\prime})^2 +m^2] - g z^{\mu}\prime A_{\mu} (z) \}
\label{eq:prop}
\end{equation}

 In place of (\ref{eq:path1})
 we find
\begin{eqnarray}
& &
G_4(x_1,x_2; y_1, y_2) =({-i\over 2})^2 \int_0^{\infty} d\tau_1
\int_0^{\infty} d\tau_2 \int_{z_1(0)=y}^{z_1(t_1)=x_1} {\cal
 D}[z_1] \int_{z_2(0)=y_2}^{z_2(\tau_2)=x_2} {\cal D} [z_2]
\nonumber \\
& & {\rm exp} {-i \over 2} \{ \int_0^{\tau_1} d \tau_1' [
 ({d z_1\over d\tau_1'})^2 + m_1^2 ] + \int_0^{\tau_2} d \tau_2'
 [ ({d z_2\over  d\tau_2'})^2 +m_2^2 ]\}
\cdot {1\over 3} \langle  {\rm Tr}\, {\rm P}\,
 {\rm exp}\, ig \oint d z^{\mu} A_{\mu} (z) \rangle
\nonumber \\
& & \quad \quad \quad
\label{eq:g4bs}
\end{eqnarray}
where  the equation of a path
 connecting $y$ with $x$ is written as
$ z^{\mu} = z^{\mu}(\tau) $,
 in terms of an arbitrary parameter
 $\tau$ (rather than   the time $t$)
 and $z'$ stands for $z(\tau')$.
Notice the occurrence again  in (\ref{eq:g4bs}) of the Wilson loop
 integral $W_{q \bar{q}}$. Even in this case we can use the
 evaluation  of $ i \ln W_{q \bar{q}}^{\rm SR} $ as given by Eq.
 (\ref{eq:cinque}) or (\ref{eq:sex})
 and assume  for  $i \ln W_{q \bar{q}}^{\rm LR}$
 the first line of (\ref{eq:dic}) in the center--of--mass system.\par
Then, by appropriate manipulations of the resulting expression,
 one can obtain the Bethe--Salpeter equation
\begin{eqnarray}
& & G_4(x_1, x_2; y_1, y_2)  =  G_2(x_1-y_1)\,  G_2(x_2-y_2)
 + \int  d^4 \xi_1 d^4 \xi_2 d^4 \eta_1 d^4 \eta_2 \nonumber \\
& & G_2(x_1-\xi)\, G_2(x_2-\xi_2)\, I(\xi_1 ,\xi_2; \eta_1 ,\eta_2)
 \, G_4 (\eta_1,\eta_2; y_1, y_2),
\label{eq:bseq}
\end{eqnarray}
with a kernel of the form
$ I(\xi_1, \xi_2;\eta_1, \eta_2)= I^{\rm SR} (\xi_1, \xi_2;
 \eta_1, \eta_2) + I^{\rm LR}(\xi_1, \xi_2;
 \eta_1, \eta_2)$.
Here $I^{\rm SR}$ coincides with  the ordinary  perturbative
 kernel, while in the momentum representation
 $I^{\rm LR}$  can be written as  (for simplicity we have
 neglected the perimeter term)
\begin{equation}
\tilde{I}^{\rm LR} (p_1', p_2'; p_1,p_2)  =
 {1\over (2 \pi)^3} \int d^3 {\bf r} e^{i ({\bf k}'- {\bf k})\cdot
 {\bf r}}  J({\bf r}, { { p}^{'}_1+p_1 \over 2},
 { p_2'+p_2 \over 2})
\label{eq:bsform}
\end{equation}
($p_1\prime +p_2\prime= p_1 +p_2 ,\,  {\bf p}_1= - {\bf p}_2= {\bf k}$,
 $ {\bf p}_1\prime= - {\bf p}_2\prime={\bf k}\prime $) with
\begin{eqnarray}
J({\bf r}, q_1, q_2)& = & (2 \pi)^3 {\sigma r \over 2}
 {1\over q_{10} +q_{20}} [ q_{20}^2 \sqrt{ q_{10}^2 -{\bf q}_{\rm T}^2}
+ q_{10}^2  \sqrt{q_{20}^2 -{\bf q}^2_{\rm T}} +\nonumber \\
& +& { q_{10}^2 q_{20}^2 \over \vert {\bf q}_{\rm T}\vert }
 ( {\rm arcsin }{ \vert {\bf q}_{\rm T} \vert \over
 \vert q_{10}\vert } + {\rm arcsin } { \vert {\bf q}_{\rm T}\vert
 \over \vert q_{20}\vert } )] + O({\sigma^2 \over m^4})
\label{eq:kerndef}
\end{eqnarray}
(having set ${\bf q}_1= -{\bf q}_2={\bf q}$, $ q^h_{\rm T}=
 (\delta^{hk} -\hat{r}^h \hat{r}^k ) q^k$).\par
Notice that, according to a standard procedure, the BS kernel
 $\tilde{I}$  can be associated  with a relativistic potential
 (to be used in the Salpeter equation) given by
\begin{equation}
\langle {\bf k}' \vert V \vert {\bf k} \rangle ={1\over
 (2 \pi)^3} { m_1 m_2 \over \sqrt{ w_1({\bf k}) w_2({\bf k})
 w_1({\bf k}') w_2({\bf k}')}} \tilde{I}_{\rm inst} ({\bf k}', {\bf k})
\end{equation}
where $w_j({\bf k})= \sqrt{m_j^2+{\bf k}^2}$
 and the instantaneous kernel $ \tilde{I}_{\rm inst}$ is obtained
 from $\tilde{I}$ by setting $p_{10}, p_{20}$ and $ p_{10}', p_{20}'$
 equal to appropriate functions of ${\bf k}$ and ${\bf k}'$,
 respectively. In the present case it is convenient to choose
 $ p_{10}=p_{10}'=\sqrt{w_1({\bf k}) w_2({\bf k}')}$ ,
 $p_{20}=p_{20}'=\sqrt{w_1({\bf k}) w_2({\bf k}')}$. If we do so
 we reobtain the Hamiltonian (\ref{eq:delta}).\par
Going back to the expression of the  kernel  $\tilde{I}^{\rm LR}$
  as given by (\ref{eq:kerndef})--(\ref{eq:bsform}),
 one has to notice that this is
 highly  singular  for $ {\bf k}'= {\bf k}$,
 due to the occurrence of the factor
 $ r$ in  (\ref{eq:kerndef}), and it must be appropriately regularized
  before being used in equation (\ref{eq:bseq}) (e.g., one can make
 the substitution $ r\to r e^{-\varepsilon r}$).
 This circumstance  is related to the fact that,
 being confining, $\tilde{I}^{\rm LR}$
   should admit only bound states, while the
 inhomogeneous BS equation provides also a continuous  two--particles spectrum.
 Therefore one should solve (\ref{eq:bseq}) for the regularized
 kernel and only  at the end take the limit
 for $\varepsilon \to 0 $ (we admit
 in this way that  resonances would evolve in bound states).

\section{Acknowledgements}
We acknowledge warmly  Prof. E.Montaldi and Dr. P.Consoli
 for having contributed
 to some of the results reported in this paper.

 \end{document}